\documentclass[twocolumn,showpacs,amsmath,amssymb,superscriptaddress]{revtex4-1}

\usepackage{dcolumn}
\usepackage{bm}
\usepackage{ifpdf}
\usepackage{hyperref}
\usepackage{dcolumn}
\usepackage{bm}
\usepackage[spanish,english]{babel}
\usepackage{amsfonts}
\usepackage{amssymb}
\usepackage{graphicx}
\usepackage[latin1]{inputenc}

\begin{document}

\title{Deforming solitons in generalized Abelian Higgs models}

\author{C. dos Santos}
\affiliation{Centro de F\'{\i}sica e Departamento de F\'{\i}sica e Astronomia, Faculdade de Ci\^{e}ncias da Universidade do Porto, 4169-007 Porto, Portugal}
\author{D. Rubiera-Garcia}
\affiliation{Departamento de F\'{\i}sica, Universidad de Oviedo, Avenida Calvo Sotelo 18, 33007 Oviedo, Asturias, Spain}

\date{\today}

\begin{abstract}
This work deals with several aspects of the extension to Abelian Higgs models of the deformation method originally developed for scalar field models. We present several examples allowing to transform self-dual solutions of different generalized Abelian Higgs models into scalar field models with or without a gauge field component. This is done through a parametrization of the soliton orbit in terms of the sine-Gordon static
kink. We extend these ideas to a nonAbelian Higgs model.

\end{abstract}


\maketitle

\section{Introduction}

The idea of the deformation method described in Ref.\cite{Bazeia0} is to find static defect solutions for a model whose equations of motion map those for another model where static defect solutions are known. The mapping is done through deformation functions that take the latter solutions into the former ones (dubbed as deformed defects).
This method can be applied several times so that one can construct an uncountable number of deformed solutions. In addition it allows to deform topological and non-topological defects, controlling their energy and width,
and generating a large amount of new static defect solutions having distinct characteristics from the original ones.

This method has been used in wide variety of contexts but always for static analytical soliton solutions, either with one or two scalar fields
\cite{Bazeia0,Bazeia1,Bazeia2,Bazeia3,Bazeia4,Bazeia5,Bazeia6,Bazeia7,Bazeia8}, while the extension to the gauge field case has not been performed yet. The main aim of this letter is to provide some preliminary insights at this regard, by considering the simplest extension of the deformation method \cite{Bazeia0}. For this purpose we consider models with a scalar and a gauge field supporting static analytical soliton solutions. For simplicity we take the self-dual ansatz by using the Bogomol'nyi-Prasad-Sommerfield (BPS) \cite{Prasad} first order reduction method. By this way (and after substituting the ansatz) our deformation problem will be reduced to that of deforming a set of two equations of motion with two scalar fields.

Now three problems arise. First, how to find the deformation functions as they should depend on both deformed fields. This means that in general one cannot use the method developed by Afonso et. al in \cite{Bazeia4} as it stands for deformation functions that only depends on the respective deformed field. Second, how to guarantee that the deformed equations of motion admit stable soliton solutions with two fields. And finally how to identify the
deformed model, a task which has non unique solutions.

The original solutions, i.e., before the deformation, are described in Ref.\cite{Bazeia4} and also in Refs. \cite{Jackiw90,Santos10,Torres97,Santos11} which correspond to solitons that have respectively two scalar fields or one single scalar field and a gauge one. They were conveniently chosen to fail the standard orbit deformation method for two scalar systems presented in Ref.\cite{Bazeia4}. Let us stress that such a method only works for two coupled equations provided that each deformation function depends only of a single independent field. Therefore either we have to find another method or/and the deformed model necessarily must have gauge fields.

The procedure developed in this paper is based on a parametrization of the soliton orbit in terms of a static kink solution of a single scalar field lagrangian to which the standard deformation method \cite{Bazeia0} can be immediately applied. By convenience we shall choose the static kink solution of the sine-Gordon model \cite{SG} as this orbit parameter, using the deformation method \cite{Bazeia0} when necessary. The chosen starting models are of two types according to whether they include or not a gauge field. Those including a gauge field are extensions of the Abelian-Higgs systems that support analytical BPS solitons with a single gauge field and a scalar one for which one cannot apply the standard orbit method deformation for two scalar fields.

We show how this procedure is able to change the topological nature and physical properties of the solitons such as their electric charge, momenta, energy density and width, depending on the model which is considered. Finally we extend these ideas by considering the nonAbelian t'Hooft-Polyakov monopole solution \cite{tHooft}.

\section{Starting models} \label{sub:I}

\subsection{Two scalar fields model}

\begin{figure}[h]
\begin{center}
\includegraphics[width=6cm,height=3cm]{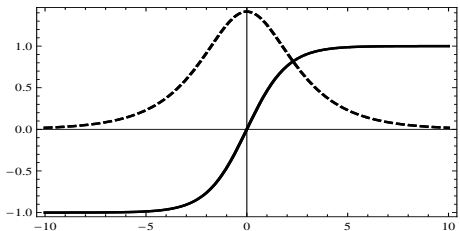}
\caption{Profile of the functions $\chi_1(x)$ (solid line) and $\chi_2(x)$ (dashed) with $r=1/4$ as given by Eq.(\ref{paramet-2scalars}) and with $\theta$ the sG field of Eq.(\ref{SG-field}). Note that these profiles reproduce the original ones given by Eq.(\ref{sol-2scalars}).}\label{fig:4}
\end{center}
\end{figure}

The first model considered here has no gauge fields and is described by the ($1+1$)-dimensional lagrangian density given by

\begin{eqnarray}
{\cal L}_{\mathcal{A}}&=&\frac{1}{2}\partial_{\mu} \chi_{1} \partial^{\mu} \chi_{1}+
\frac{1}{2}\partial_{\mu} \chi_{2} \partial^{\mu} \chi_{2}- \nonumber \\ &-& \frac{1}{2}\left(1-\chi_{1}^2-r \chi_{2}^2\right)^2 +2r^2 \chi_{1}^2 \chi_{2}^2,
\end{eqnarray}
with $r$ a real parameter. The (static) equations of motion are
\begin{equation} \label{def-2scalars}
\chi_{1}^{\prime}=1-\chi_{1}^2-r \chi_{2}^2 \hspace{0.2cm}; \hspace{0.2cm} \chi_{2}^{\prime}=-2r \chi_{1} \chi_{2}.
\end{equation}
where $'$ stands for derivative with respect to $x$. Taking $0<r\leq 1/2$ there is a soliton solution \cite{Rajaraman}
\begin{equation} \label{sol-2scalars}
\chi_{1}=\tanh(2rx)\hspace{0.2cm}; \hspace{0.2cm}  \chi_{2}=\sqrt{\frac{1-2r}{r}}\text{sech}(2rx),
\end{equation}
verifying the soliton orbit
\begin{equation} \label{orb-2scalars}
\left(\sqrt{\frac{r}{1-2r}} \,\,\chi_{2}\right)^2+\chi_{1}^2=1,
\end{equation}
which is the equation of a circle, to which the standard two scalar fields deformation method \cite{Bazeia4} can be applied.
Taking the following parametrization
\begin{equation} \label{paramet-2scalars}
\sqrt{\frac{r}{1-2r}} \,\,\chi_{2}=\sin(\theta)\hspace{0.2cm}; \hspace{0.2cm} \chi_{1}=-\cos(\theta),
\end{equation}
the equations (\ref{def-2scalars}) give the sine-Gordon (sG) equation of motion (see Eqs.(\ref{sinegordon}), (\ref{SG-field}) below)
\begin{equation}\label{tau-sine-gordon}
\frac{d\theta}{d\tau}=\sin(\theta),
\end{equation}
where $\tau = 2rx$. In Fig.\ref{fig:4} we show that the plots of $h(x)$ and $A(x)$.

In the next two sections we select two kinds of Abelian-Higgs models supporting soliton orbits to which the standard two scalar fields orbit deformation method cannot be generically applied. We present the soliton solutions and the respective orbit. We also present our procedure used for deformation which is based on a parametrization of the soliton fields in terms of the sine-Gordon static kink.

\subsection{$w$ models}

These models are extensions of the Abelian-Higgs one and are described in Refs.\cite{Santos10,Santos11}, which we now briefly review. They
are defined in ($2+1$) dimensions where the Maxwell term is replaced by a Chern-Simons one and the kinetic term is non-canonical, as described by the lagrangian density
\begin{equation}
{\cal L}_{w}=\omega (| \varphi |)X-V(|\varphi|)+\frac{k}{4}\epsilon^{\alpha \beta \gamma} A_{\alpha} F_{\beta \gamma}
\end{equation}
where $k$ is a constant, $\omega(| \varphi |)$ is a function of the complex scalar (Higgs) field $\varphi$, $X=| D_{\mu} \varphi |^2$ is the standard canonical kinetic term, $V(|\varphi|)$ is the potential, assumed to implement a symmetry breaking mechanism, $A^{\mu}=(A^0,\textbf{A})$ is the (abelian) gauge field and $F_{\mu\nu}=\partial_{\mu}A_{\nu}-\partial_{\nu}A_{\mu}$ is the field strength tensor. The associated electric and magnetic fields are defined in the usual way, i.e. $\textbf{E}^i=F^{i0}=-\frac{d\textbf{A}}{dt}-\nabla_i A^0$ and $\textbf{B}=\vec{\nabla} \times \textbf{A}$, where we are using bold text style as the notation for vectors.

For static solitons with axial symmetry along the $y$ axis we take the ansatz
\begin{equation}\label{ansatz1}
\varphi  = \frac{e}{\sqrt{k}}\,h(x) \hspace{0.2cm}; \hspace{0.2cm}  \mathbf{A}=\frac{e}{k}(0, A(x)),
\end{equation}
with $A_0^2 = A_y^2$ and $h(x)$ real and thus the equations of motion for BPS solutions (see e.g.\cite{Santos10} for details on this procedure) become
\begin{equation}\label{eom-walls}
h'=\pm hA\hspace{0.2cm}; \hspace{0.2cm} A'=-2\omega h^2 A_0,
\end{equation}
with $A_0$ given by the Gauss law constraint $A_0=\frac{B}{2\,h^2\,w}$.
Let us stress that in these models the potential depends on the choice of $\omega(h)$ as a consequence of the equations (\ref{eom-walls}).

In the next sections we take particular choices for $\omega(h)$ leading us back to several models already considered in the literature.

\subsubsection{Jackiw's wall and sine-Gordon kink} \label{sub:II}

\begin{figure}[h]
\begin{center}
\includegraphics[width=6cm,height=3cm]{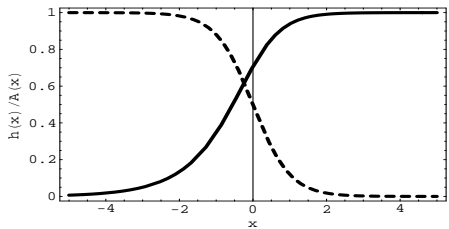}
\caption{Profile of the functions $h(x)$ (solid) and $A(x)$ (dashed) through the parametrization (\ref{orbit-jackiw}), with $\psi$ given in (\ref{SG-field}). Note that these profiles reproduce those of Jackiw's solution (\ref{sol-jackiw}).}\label{fig:1}
\end{center}
\end{figure}

For $\omega=1$ the equations of motion (\ref{eom-walls}) admit a wall-like soliton solution, the so-called Jackiw's wall \cite{Jackiw90} described by

\begin{equation} \label{sol-jackiw}
h= \frac{1}{\sqrt{1+e^{-2x}}} \hspace{0.2cm}; \hspace{0.2cm}  A= \frac{1}{1+e^{2x}}.
\end{equation}
verifying the soliton orbit
\begin{equation}\label{orb-Jackiw}
\left(\sqrt{A}\right)^2+h^2=1.
\end{equation}
which is the equation of a circle. In this case the standard orbit method deformation for two scalar fields can be applied in order to make the deformation. In fact, it is possible to apply that method by taking $\chi_1$ and $\chi_2$ the deformed fields as $\chi_1=\sqrt{A}$ and $\chi_2=h$. However the deformed potential does not support solitons. Thus, we proceed developing our method of parameterizing the soliton fields in terms of the sine-Gordon one. For that we note that if we take the parametrization
\begin{equation} \label{orbit-jackiw}
\sqrt{A}=\cos\left(\frac{\psi}{2}\right) \hspace{0.2cm}; \hspace{0.2cm}  h=\sin\left(\frac{\psi}{2}\right).
\end{equation}
the equations (\ref{eom-walls}) are reduced to
a single one
\begin{equation} \label{sinegordon}
\psi'=\sin(\psi),
\end{equation}
which is precisely the sine-Gordon equation. Thus $\psi$ corresponds to the static sG kink, solution of (\ref{sinegordon}), which is given by
\begin{equation}\label{SG-field}
\psi= 2\arctan[e^{x}],
\end{equation}
In this way the Jackiw's wall is parameterized in terms of the sG field. In Fig.\ref{fig:1} $h(x)$ and $A(x)$ are plotted by using Eqs.(\ref{orbit-jackiw}) with $\psi$ given by Eq.(\ref{SG-field}).

\subsubsection{Generalized and sine-Gordon kinks} \label{sub:II}

Let us now take the non-canonical model studied in Ref.\cite{Santos10} and characterized by the function
\begin{equation}
w(h) = \frac{2}{(1+h^2)^2}.
\end{equation}
The equations of motion (\ref{eom-walls}) now admit a wall-like soliton given by
\begin{equation} \label{sol-carol}
h(x)=\sqrt{1+e^{-2x}}-e^{-x} \hspace{0.2cm}; \hspace{0.2cm}  A(x)=\frac{1}{\sqrt{1+e^{2x}}},
\end{equation}
which verifies the hyperbolic soliton orbit
\begin{equation} \label{orb-carol}
\left(h+\frac{A}{\sqrt{1-A^2}}\right)^2-\frac{A^2}{1-A^2}=1.
\end{equation}
This is an example to which the standard orbit method deformation for two scalar fields cannot be applied. Had we applied it, the deformed fields
$\chi_1$ and $\chi_2$, would be $|\chi_1|=h+\frac{A}{\sqrt{1-A^2}}$ and $|\chi_2|=\frac{A}{\sqrt{1-A^2}}$. However its equations of motion cannot be derived from a BPS scalar system as an inconsistency when writing the superpotential emerges.

So, let us apply our parametrization method followed by the standard deformation one \cite{Bazeia0}. A suitable parametrization for the orbit (\ref{orb-carol}) is given by

\begin{equation} \label{orbit-carol}
h+\frac{A}{\sqrt{1-A^2}}=\cosh (\theta)\hspace{0.2cm}; \hspace{0.2cm} \frac{A}{\sqrt{1-A^2}}= - \sinh (\theta).
\end{equation}
By making some manipulations and constraining $\theta \leq 0$, one gets that $h$ and $A$ are given by

\begin{equation}
h= \cosh (\theta) + \sinh (\theta)\hspace{0.2cm}; \hspace{0.2cm}  A= - \tanh(\theta).
\end{equation}
Consistency with the system (\ref{eom-walls}) requires that $\theta'=-\tanh(\theta)$, which can be explicitly integrated to give $\theta=arcsinh[e^{-x}]$. This is the equation of motion of a scalar field theory with lagrangian given by

\begin{equation} \label{scalar-3}
L=\frac{(\theta')^2}{2}-\frac{1}{2}\tanh^2(\theta)
\end{equation}
However, we want to relate this model to the sG one as this will allow us later to transform BPS solutions of different Abelian Higgs models between themselves. For this purpose we now employ the standard deformation method \cite{Bazeia0} defining a new scalar field $\psi$ and lagrangian density as

\begin{equation}
L(\psi)=\frac{1}{2}(\psi')^2-\frac{1}{2}\tanh^2 (\theta)\left(\frac{d \theta}{d \psi}\right)^{-2},
\end{equation}
and requiring the new potential to be the sG one (and thus $\psi$ the sG field). Performing an integration we are led to the relation $\log[\tan(\psi/2)]=\pm \log[|\sinh(\theta)|]$ between both scalar fields. It proves that the sign ($-$) in this relation is the one that leads
to the sG solution. Therefore we obtain $h$ and $A$ as functions of the sG field $\psi$ as

\begin{equation} \label{dualwall}
h= \frac{1-\cos[\psi/2]}{\sin[\psi/2]}\hspace{0.2cm}; \hspace{0.2cm}  A= \cos[\psi/2].
\end{equation}

\begin{figure}[h]
\begin{center}
\includegraphics[width=6cm,height=3cm]{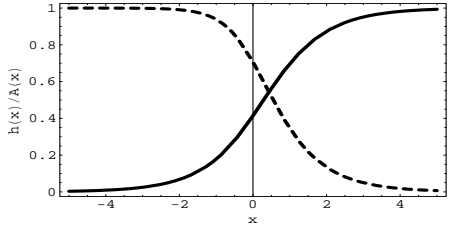}
\caption{Profile of the functions $h(x)$ (solid line) and $A(x)$ (dashed) as given by Eq.(\ref{dualwall}) with $\psi$ the sG field of Eq.(\ref{SG-field}). Note that these profiles reproduce the original ones given in (\ref{sol-carol}).}\label{fig:2}
\end{center}
\end{figure}

\subsubsection{Generalized lump and sine-Gordon kink} \label{sub:IV}

Let us consider another model supporting soliton solutions for which the orbit deformation method fails. It has been studied in Ref.\cite{Santos11} and it is defined by the choice

\begin{equation} \label{lump}
\omega(h)=\frac{h^2}{2\sqrt{1-h^4}},
\end{equation}
Solving the BPS field equations (\ref{eom-walls}) we obtain the soliton fields

\begin{equation} \label{sol-lump}
h(x)=(\cosh(x))^{-1/2}\hspace{0.2cm}; \hspace{0.2cm} A(x)=-\frac{1}{2} \tanh(x).
\end{equation}
satisfying the soliton orbit $(2A)^2+(h^2)^2=1$ which is again a circle. Had we applied the orbit deformation method then the deformed fields,
$\chi_1$ and $\chi_2$, would be such that $|\chi_1|=2A$ and $|\chi_2|=h^2$ and we would arrive again to an inconsistency with the superpotential when trying to derive the BPS equations of motion. Instead let us introduce the following parametrization

\begin{equation} \label{lumpsol}
h=\sin^{1/2}(\psi)\hspace{0.2cm}; \hspace{0.2cm}  A=\frac{1}{2}\cos(\psi),
\end{equation}
so the consistency with (\ref{eom-walls}) requires that $\psi'=\sin(\psi)$ which is the sG equation and thus $\psi$ the sG field. Therefore no further deformation is needed and Eq.(\ref{lumpsol}) performs the transformation between the BPS solutions of the Abelian Higgs system in \cite{Santos11} and the sG kink.

\subsection{Torres model} \label{sub:V}

Finally let us consider another gauge model supporting solitons that cannot be deformed through the orbit deformation method.
The model is the one described in Ref.\cite{Torres97}, which is a scalar Quantum Electrodynamics model in ($2+1$) dimensions with the addition of the Chern-Simons term and an anomalous magnetic interaction, described by the lagrangian density

\begin{equation}
{\cal L}_{\mathcal{T}}=-\frac{1}{4} F_{\mu\nu}F^{\mu\nu} + \frac{k}{4}\epsilon^{\alpha \beta \gamma} A_{\alpha} F_{\beta \gamma} + \frac{1}{2}  |{\cal{D}}_\mu \phi |^2-\frac{1}{2}m^2 |\phi|^2
\end{equation}
where ${\cal{D}}_\mu = \partial_\mu - ieA_\mu-i\frac{g}{4}\epsilon_{\mu\nu\alpha}F^{\nu\alpha}$ is the gauge-covariant derivative and $g$ an anomalous magnetic moment. For $g=-\frac{2e}{k}$ the equations of motion reduce to first order and for the ansatz $\phi =(k/e)f$ with $A_\mu=(0,0,A_y)$ they become
\begin{equation} \label{eq-torres}
f'=\mp mf \hspace{0.2cm}; \hspace{0.2cm}   A_y'= \mp \frac{kf^2}{(1-f^2)}A_y.
\end{equation}

\begin{figure}[h]
\begin{center}
\includegraphics[width=6cm,height=3cm]{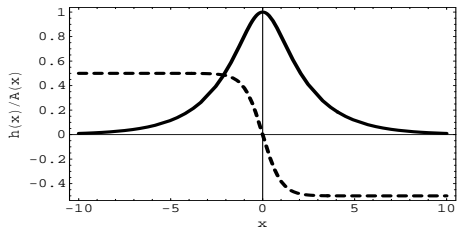}
\caption{Profile of the functions $f(x)$ (solid line; taking $m=1$) and $A_y(x)$ (dashed; $m=1, \gamma=1,k=2$) centered at $X=0$, as given by Eq.(\ref{torressol}) and $\psi$ is the sG field of Eq.(\ref{SG-field}). Note these profiles reproduce the original ones given by Torres (\ref{sol-torres}).}\label{fig:3}
\end{center}
\end{figure}

Eqs.(\ref{eq-torres}) admit the soliton solution

\begin{eqnarray} \label{sol-torres}
f&=& e^{-m| x-X |} \nonumber \\
A_y&=& Sign(x-X)\frac{\gamma}{2}\left(1-e^{-2m|x-X|}\right)^{\frac{k}{2m}}.
\end{eqnarray}
which is a ``spiked" lump-like soliton with symmetry along the $y$ axis, located at $x=X$
with width of order $1/m$, carrying a magnetic flux per unit of length of $\gamma$ and having
a charge per unit of length of $-k\gamma$. The profiles for $f(x)$ and $A_y(x)$ are given in Fig.\ref{fig:3}.

The soliton orbit satisfied by these solutions is obtained as

\begin{equation} \label{orbit-torres}
\left[\left(\frac{A_y}{Sign(x-X)\frac{\gamma}{2}}\right)^{\frac{m}{k}}\right]^2+f^2=1.
\end{equation}
which is again the equation of a circle. As before, applying the deformation orbit method the deformed fields, $\chi_1$ and $\chi_2$, would be such that $|\chi_1|=\frac{A_y}{Sign(x-X)\frac{\gamma}{2}}$ and $|\chi_2|=f$ with the same result for the consistence of the superpotential. Proceeding in the same way as in the previous examples, take the parametrization

\begin{equation} \label{par-torres}
\left(\frac{A_y}{Sign(x-X)\frac{\gamma}{2}}\right)^{\frac{m}{k}}=\cos(\theta)\hspace{0.2cm}; \hspace{0.2cm} f =\sin(\theta),
\end{equation}
and the consistency with the equations (\ref{eq-torres}) brings that $\theta'=\mp m \tan(\theta)$. These correspond to the equation of motion for the lagrangian

\begin{equation}
L=\frac{1}{2}\partial_{\mu}\theta \partial^{\mu}\theta -\frac{1}{2}m^2 \tan^2(\theta).
\end{equation}
In order to relate the original Torres model to the sG one we use the standard deformation method for a single field to get the deformed lagrangian density as

\begin{equation} \label{lag-torres-scalar}
L(\psi)=\frac{1}{2}\left(\psi'\right)^2-\frac{1}{2}m^2 \tan^2(\theta)\left(\frac{d \theta}{d \psi}\right)^{-2},
\end{equation}
where the second right-hand-side term is identified as the potential for the sG solution, i.e.

\begin{equation} \label{def-torres}
\frac{1}{2}\sin^2 (\psi)=\frac{1}{2}m^2 \tan^2(\theta) \left(\frac{d \theta}{d \psi}\right)^{-2}.
\end{equation}
Solving this equation we obtain that $m \log[\tan(\psi/2)]=\pm \log[\sin(\theta)]$ where to obtain the sG system only the sign $(-)$ must be taken. Thus one is led to the relation $|\sin(\theta)|=[cotg(\psi/2)]^m$, which introduced into Eq.(\ref{par-torres}) gives that
\begin{eqnarray} \label{torressol}
f&=&cotg^m[\psi/2]\nonumber \\
A_y&=&Sign(x-X) \frac{\gamma}{2}\left(1-\cot^{2m}[\psi/2]\right)^{\frac{k}{2m}}. \nonumber
\end{eqnarray}


\section{Transforming the BPS solutions} \label{sec:transformations}

In the previous sections we related different BPS soliton solutions of several Abelian gauge field models to the same
soliton solution (the sG one). This makes possible to relate them between themselves.
In order to do this, we shall consider five different cases, which are the combination of transforming
equal/different kinds of solitons in equal/different theories, with an additional case for the two-scalar fields system.

\subsection{Solitons of the same kind in the same theory} \label{A}

\begin{figure}[h]
\begin{center}
\includegraphics[width=8.5cm,height=3cm]{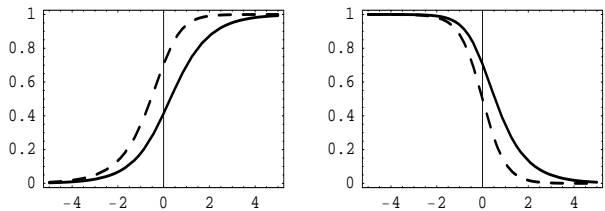}
\caption{Left panel: $h(x)$ profile for Jackiw's wall (dashed) and the generalized wall (solid). Right panel: The same for $A(x)$.}\label{fig:5}
\end{center}
\end{figure}

\begin{figure}[h]
\begin{center}
\includegraphics[width=7cm,height=3.cm]{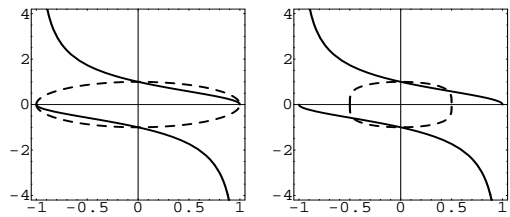}
\caption{Orbits before (dashed) and after (solid) the transformation for the case \ref{A} (left panel) and \ref{B} (right panel).}\label{fig:6}
\end{center}
\end{figure}

Let us begin by transforming the Jackiw's wall (\ref{sol-jackiw}) into the generalized BPS wall (\ref{sol-carol}). Labeling by $h_J$, $A_J$ the original fields and by $\phi_1$, $\phi_2$ the ``deformed" fields, we simply need Eqs.(\ref{orbit-jackiw}) and (\ref{dualwall}) to conclude that

\begin{equation}
\phi_1(h_J,A_J)=\frac{1-\sqrt{A_J}}{\sqrt{h_J}}\hspace{0.2cm}; \hspace{0.2cm}  \phi_2=\sqrt{A_J}
\end{equation}
gives the transformation functions. In Fig.\ref{fig:5} we compare the fields before and after the transformation. Note that the only physical difference between the deformed and non-deformed soliton lies on their thickness. The orbits before and after the transformation are also shown in Fig.\ref{fig:6}.

\subsection{Solitons of different kind in the same theory} \label{B}

Now we transform the generalized BPS wall (\ref{dualwall}) (fields $h, A$) into the lump (\ref{lumpsol}) (fields $\phi_1,\phi_2$). Simple trigonometric relations give that

\begin{eqnarray}
\phi_1(h_W,A_W)&=& \pm \sqrt{\frac{2A_W(1-A_W)}{h_W}} \nonumber \\
\phi_2(h_W,A_W)&=&A_W^2-\frac{1}{2}.
\end{eqnarray}

Noting that the linear total momentum along the $y$ axis depends on the difference $A^2(\infty)-A^2(-\infty)$ (see Refs.\cite{Santos10,Santos11} for further details) and using Fig.\ref{fig:5} one can see that the only physical difference between the starting and resulting solitons lies on this physical quantity which vanishes for the lump. The orbits before and after this transformation are also shown in Fig.\ref{fig:6}.

\subsection{Solitons of the same kind in different theories, both with gauge fields} \label{C}

Let us now consider the transformation of the lump-like solution of Eq.(\ref{lumpsol}) into Torres lump (\ref{torressol}). Using trigonometric relations this is easily achieved with the result

\begin{eqnarray}
f(h_L,A_L)&=& \left(\frac{2A_L+1}{h_L^2}\right)^{\frac{m}{2}} \\
A_y(h_L,A_L)&=& Sign(x-X)\frac{\gamma}{2} \left(1-\left[\frac{2A_L+1}{h_L^2}\right]^{2m}\right)^{\frac{k}{2m}} \nonumber
\end{eqnarray}

\begin{figure}[h]
\begin{center}
\includegraphics[width=7cm,height=3.0cm]{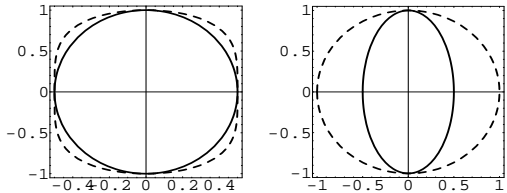}
\caption{Orbits before (dashed) and after (solid) the transformation for the case \ref{C} (left panel) and \ref{D} (right panel).}\label{fig:7}
\end{center}
\end{figure}

The only physical difference between both solitons lies on their interaction with an external magnetic field as
the lump of Torres has a non-vanishing magnetic moment. In Fig.\ref{fig:7} we plot the original and final orbits.

\subsection{Solitons of different kind in different theories, both with gauge fields} \label{D}

We consider the transformation of Jackiw's wall (\ref{sol-jackiw}) into Torres lump (\ref{torressol}). In this case no further trigonometric manipulations are needed and thus we immediately obtain

\begin{eqnarray}
f(h_J,A_J)&=&\left(\frac{\sqrt{A_J}}{h_J}\right)^m \\
A_y(h_J,A_J)&=&\pm Sign(x-X) \frac{\gamma}{2} \left(1-\left(\frac{\sqrt{A_J}}{h_J}\right)^{2m}\right)^{\frac{k}{2m}} \nonumber
\end{eqnarray}

The physical difference between both soliton lies on their linear total momentum along the $y$ axis which vanishes for the lump and also on its interaction with an external magnetic field as the lump of Torres has a non vanishing magnetic moment. In Fig.\ref{fig:7} we plot the orbits associated to the starting and resulting models.

\subsection{Solitons of different kind in different theories, one without gauge field} \label{E}

\begin{figure}[h]
\begin{center}
\includegraphics[width=6cm,height=3.5cm]{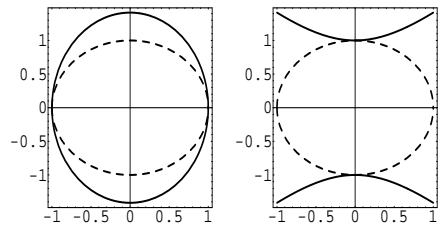}
\caption{\label{fig:epsart} Orbits before (dashed) and after (solid) the transformation for the cases \ref{E} (left panel) and \ref{F} (right panel). Note that in the solid orbit in the right panel we are plotting in the axis $u-v$.}\label{fig:8}
\end{center}
\end{figure}

We finally consider the transformation of Jackiw's wall (\ref{sol-jackiw}) into a soliton system of two scalar fields
(\ref{sol-2scalars}). Through Eqs.(\ref{orbit-jackiw}) and (\ref{paramet-2scalars}) one obtains

\begin{equation}
\chi_1(h_J,A_J)=1-2A_J\hspace{0.25cm}\chi_2=2\sqrt{\frac{1-2r}{r}}h_J\sqrt{1-h^2_J}
\end{equation}
as the functions connecting these solutions. Note that there are two ways to write $\chi_1$ and $\chi_2$ in function of $h_J$ and $A_J$, i.e. $h_J$ with $\chi_1$ and $A_J$ with $\chi_2$, as the original/deformed pairs, or the opposite. The resulting two fields soliton has neither electric charge nor magnetic flux density nor linear total momentum, as opposed to the starting soliton. The orbits for these two cases are shown in Fig.\ref{fig:8}.

\section{Extension to nonAbelian gauge fields} \label{F}

The procedure underlined above may be extended to more complex systems. As an example let us consider the nonAbelian Higgs model described by the lagrangian

\begin{equation}
L=D_{\mu}\Phi^aD^{\mu}\Phi^a-\frac{1}{4}F_{\mu\nu}^a F^{\mu\nu a}-V(\Phi^a),
\end{equation}
where $D_{\mu}\Phi^a=\partial_{\mu}\Phi^a-e_0 \varepsilon^{abc}A_{\mu}^b \Phi^c$ is the gauge-covariant derivative and $V(\Phi^a)=\frac{\lambda}{4}(\Phi^a \Phi^a -\eta^2)^2$ is the standard Higgs potential. After implementing the \emph{hedgehog} ansatz

\begin{equation}
\Phi^a = H(x) \frac{x^a}{r} \hspace{0.1cm};\hspace{0.1cm}  A_0^a = 0 \hspace{0.1cm};\hspace{0.1cm} A_i^a=\varepsilon_{iak} \frac{x_k}{e_0 r^2} [ \Omega(x)-1],
\end{equation}
and taking the limit $\eta \rightarrow 0$ with $\eta=\frac{\sqrt{2\lambda}}{e_0}$, the equations of motion become of first order and given by

\begin{equation}
H'=\frac{1-\Omega^2}{r^2} \hspace{0.2cm};\hspace{0.2cm} \Omega'=-\Omega H.
\end{equation}
Its solution is the t' Hooft-Polyakov monopole \cite{tHooft}

\begin{equation} \label{monopole}
\Omega=\frac{r}{\sinh r} \hspace{0.2cm};\hspace{0.2cm} H(r)=\frac{1}{\tanh r}-\frac{1}{r}.
\end{equation}

Now let us rewrite the equations of motion by using the variables $u=H+\frac{1}{r}, v=\frac{\Omega}{r}$, which implies $u'=-v^2$ and $v'=-vu$, and by integrating we are led to the orbit $u^2-v^2=C$ with $C$ a real constant, whose value determines different orbits. Let us take $C=1$ (see Fig.\ref{fig:8} for the associated orbit), for which the t'Hooft-Polyakov monopole solution is recovered. The parametrization

\begin{equation}\label{uv}
u=-\cosh (\theta) \hspace{0.2cm};\hspace{0.2cm} v= -\sinh (\theta),
\end{equation}
leads to the sinh-Gordon equation

\begin{equation} \label{sinhG}
\theta'=\sinh (\theta),
\end{equation}
which has the analytical solution $\theta(x)=2 arctanh[e^{x}]$. The lagrangian density corresponding to Eq.(\ref{sinhG}) is given by

\begin{equation}
L=(\theta')^2-\frac{1}{2}\sinh^2 (\theta)
\end{equation}

Now we use the standard deformation method for a single scalar field to get the deformed lagrangian density as

\begin{equation}
L =  (\psi')^2-\frac{1}{2} \sinh^2 (\theta) \left(\frac{d \theta}{d \psi}\right)^{-2}
\end{equation}
with the second right-hand-side term being identified as the sG potential. This gives that
$\frac{\psi'}{\sin \psi} = + \frac{\theta'}{\sinh \theta}$ and thus

\begin{equation}
\psi=2\arctan[\tanh(\theta/2)].
\end{equation}
\begin{figure}[h]
\begin{center}
\includegraphics[width=6cm,height=3.5cm]{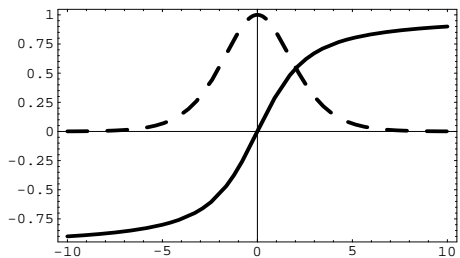}
\caption{\label{fig:epsart} Profile of the functions $H(r)$ (solid line) and $\Omega(r)$ (dashed) as given by the original t'Hooft-Polyakov solution (\ref{monopole}) and the parametrization (\ref{uv}).}\label{fig:9}
\end{center}
\end{figure}
In Fig.\ref{fig:9} we plot the $H$ and $\omega$ fields, as given by the original fields (\ref{monopole}) and the resulting ones introduced through the scalar field $\psi$. As expected both functions coincide.

As in section \ref{sec:transformations}, the hedgehog solutions of this nonAbelian model can be transformed into the BPS solutions of any of the other models. For simplicity let us consider the transformation of Jackiw's wall (\ref{sol-jackiw}) into this nonAbelian system. As both systems are written in terms of $\psi/2$ with $\psi$ the sG field, this is straightforwardly obtained as

\begin{eqnarray} \label{jacmon}
H&=&-\cosh[2arctanh[-h_J(r)/\sqrt{A_J(r)}]]-1/r \nonumber \\
\Omega &=& r \sinh[2arctanh[-h_J(r)/\sqrt{A_J(r)}]],
\end{eqnarray}
where we have relabeled the variable $x \rightarrow r$ in Jackiw's wall functions. The orbits before/after are plotted in Fig.\ref{fig:8} (right panel). This example shows how the procedure described in this paper is also suitable for nonAbelian gauge fields. \\

\section{Conclusions}

We have presented a way of relating theories with either two scalar fields or a scalar field and a single independent gauge field component (BPS solutions of Abelian Higgs models). It is based on the parametrization of the soliton orbits in terms of a scalar field. Once this is done one may use the standard deformation method of Ref.\cite{Bazeia0} to deform this scalar field into another one. This auxiliary tool allows BPS solitons of different Abelian Higgs fields to be transformed between themselves since they all can be parameterized
in terms of the same scalar field, that we take to be the sine-Gordon one. With the examples provided we give some insights not only on the extension of the deformation method to the gauge field case, but also to models where the kinetic term is non-canonical (see examples \ref{sub:II}, \ref{sub:IV} and \ref{sub:V}). The physics of the deformed solitons can be quite different from that for the original ones depending on the models involved. In the examples shown the procedure employed is able to change the width of the original soliton implementing on it an electric charge, a magnetic flux, a linear momentum or even a magnetic momentum. Moreover, through an explicit example we have shown that this idea could be applicable to models with more than one gauge field independent component, a possibility that deserves to be fully explored in a future work.

\acknowledgments

C. dS. is partially funded under the FCT project CERN/FP/116358/2010. D. R.-G. is partially funded by the Centro de F\'{\i}sica do Porto, and would like to thank the Departamento de F\'{\i}sica e Astronomia da Faculdade de Ciencias da Universidade do Porto for their hospitality. The authors are indebted to L. Losano and J. Mateos-Guilarte for useful comments and D. Bazeia for suggesting the idea of this research.

\end{document}